\documentclass{aa}  

\usepackage{graphicx}
\usepackage{xcolor}
\usepackage{mhchem}
\usepackage{txfonts}
\providecommand{\e}[1]{\ensuremath{\times 10^{#1}}}

\usepackage{hyperref}
\begin{document}

   \title{Hydrogen abstraction reactions in formic and thioformic acid isomers by hydrogen and deuterium atoms}

   \subtitle{Insights on isomerism and deuteration}

   \author{G. Molpeceres
          \inst{1}
          \and 
          I. Jim\'enez-Serra
          \inst{2}
          \and
          Y. Oba
          \inst{3}
          \and
          T. Nguyen
          \inst{3}
          N. Watanabe
          \inst{3}
          \and 
          J. Garc\'ia de la Concepci\'on
          \inst{2}
          \and
          B. Mat\'e
          \inst{4}
          \and
          R. Oliveira
          \inst{5}
          \and
          J. K\"astner
          \inst{1}
          }

   \institute{Institute for Theoretical Chemistry, University of                Stuttgart, Stuttgart, Germany\\
              \email{molpeceres@theochem.uni-stuttgart.de}
         \and
            Centro de Astrobiología (CSIC-INTA), Ctra. de Ajalvir Km. 4, Torrejón de Ardoz, 28850 Madrid, Spain\\
              \email{ijimenez@cab.inta-csic.es}
         \and
            Institute of Low Temperature Science, Hokkaido University, Sapporo, Hokkaido, Japan \\
            \email{oba@lowtem.hokudai.ac.jp}
         \and
            Insituto de Estructura de la Materia (IEM-CSIC), IEM-CSIC, Calle Serrano 121, 28006 Madrid, Spain 
         \and
            Chemistry Institute, Federal University of Rio de Janeiro, Rio de Janeiro, Brazil
             }

   \date{Received \today; accepted \today}

 
  \abstract
   {The isomerism of molecules in the interstellar medium and the mechanisms behind it are essential questions in the chemistry of organic molecules in space. In particular, for the simple formic and thioformic acids, the low temperatures found in molecular clouds indicate that cis-trans isomerization in the gas-phase must be impeded. Reactions happening on top of interstellar dust grains may explain the isomer interconversion at low temperatures.}
   {We studied the isomerization processes of formic and thioformic acid susceptible to happen on the surface of interstellar dust grains and initiated by \ce{H} abstraction reactions. Similarly, deuterium enrichment of the acids can occur by the same mechanism. Our objective is to shed light on both topics to increase our understanding of key precursors of organic molecules in space. }
   {We determined the rate constants for the \ce{H} abstraction reactions as well as the binding energies for the acids on water ice using \textit{ab-initio} calculations and the instanton method for calculating the rate constants, including quantum tunneling. In addition, we tested the viability of the deuteration of formic acid with tailored experiments and looked for it on the L1544 source. }
   {For formic acid, there is a clear dependence of the \ce{H} abstraction reactions on the isomer of the reactant, with rate constants at $\sim$50~K that differ by 5 orders of magnitude. Correspondingly, we did not observe the trans-cis reaction in our experiments. In the case of thioformic acid, a very similar cis-trans reactivity is found for abstraction reactions at the thiol (-SH) group in contrast to a preferential reactivity found when abstractions take place at the -CH moiety. We found comparable binding energies for both isomers with average binding energies of around $-6200$~K  and $-3100$~K for formic and thioformic acid, respectively. Our binding energy calculations show that the reactions are precluded for specific orientations, affecting the overall isomerization rate. For H abstractions initiated by deuterium atoms, we found very similar trends, with kinetic isotope effects varying in most cases between 13 and 20.}
   {Our results support the cis-trans interconversion of cis-formic acid on dust grains, suggesting that such an acid should not withstand the conditions found on these objects. On the other hand, the trans isomer is very resilient. Both isomers of thioformic acid are much more reactive. Non-trivial chemistry operates behind the apparent excess of its trans isomer found in cold molecular clouds and star-forming regions due to a subtle combination of preferential reactivity and binding with the surface. In the light of our results, all the deuterated counterparts of thioformic acid are viable molecules to be present on the ISM. In contrast, only the trans isomer of deuterated formic acid is expected, for which we provide upper bounds of detection. With the mechanisms presented in this paper, other mechanisms must be at play to explain the tiny fraction of cis-formic acid observed in interstellar cold environments, as well as the present trans-DCOOH and trans-HCOOD abundances in hot-corinos.  }

   \keywords{ISM: molecules -- Molecular Data -- Astrochemistry -- methods: numerical -- methods: observational -- methods: laboratory: molecular }

   \maketitle
%

\section{Introduction}

Recent detections of molecules toward different sources show that, under the harsh conditions of the interstellar medium (ISM), many molecules present different isomerism than on Earth. As an example, formic acid (FA, HC(O)OH) is mostly found (>99.9 \%) in its trans isomer under laboratory conditions \citep{Macoas2003}. In space, cis-formic acid (c-FA) is much more abundant, between 33\% measured in photodissociation regions (PDR) \citep{Cuadrado2016} to around 6\% measured in cold cores \citep{Taquet2017,Agundez2019}. In the core of the origin of isomerism in space, there is a rich interplay between different chemical and physical processes. Among those we can find: specific formation routes both in the gas phase and on grains \citep{Vazart2015,Loomis2015,Quan2016,Melli2018, Bizzocchi2020, Molpeceres2021d}, quantum-tunneling mediated isomerization \citep{GarciaConcepcion2021,delaconcepcion2021transcis}, photoisomerization \citep{Cuadrado2016} or chemical conversion after formation on grains \citep{Shingledecker2020}. 

The previously mentioned FA is a well-known molecule in interstellar chemistry, e.g., detected in pre-stellar cores, protostars, or protoplanetary discs \citep{Zuckerman1971,Cernicharo2012,Vastel2014, Cuadrado2016, Favre2018, Agundez2019, RodriguezAlmeida2021}. Similarly, thioformic acid (TFA, HC(O)SH) has been recently detected toward the giant molecular cloud G+0.693-0.027 \citep{RodriguezAlmeida2021} and tentatively in the hot molecular core G31 \citep{delaconcepcion2021transcis}. Observations indicate that the trans-thioformic acid (t-TFA) is more abundant than its cis form \citep{RodriguezAlmeida2021,delaconcepcion2021transcis}. These acids are fundamental, as the gateway of complex organic molecules (COM) \citep{Hollis2000,Modica2010, Vastel2014, JimenezSerra2016, Oberg2016}. Similarly, TFA is an essential precursor in prebiotic and primordial Earth processes \citep{Chandru2016, Foden2020}.

At the low temperatures of cold, molecular cores, molecules of the size of FA and TFA deplete on top of ice-coated interstellar dust grains. In the early stages of a molecular cloud, the ice composition of the ice mantle is amorphous solid water (ASW) \citep{Boogert2015}. Although it is estimated that around 1\% of the total mass of molecular clouds is locked on grains, rich chemistry happens on their surfaces. Reactions with atomic hydrogen are a viable source of chemical richness on the surface of dust grains. An important type of reaction of adsorbates that significantly alters the isomeric excess are \ce{H} abstraction reactions. Following a cycle of \ce{H} abstraction and subsequent addition, a particular isomer can be interconverted into the other. In addition to \ce{H} abstraction reactions, H addition reactions can also alter the isomeric excess.

In this work, we put the focus on \ce{H} abstraction reactions from FA and TFA acids on grains, continuing our recent efforts to understand their chemistry and isomerism. In addition to the formation routes of TFA \citep{Nguyen2021, Molpeceres2021d} we have very recently tackled the tunneling-mediated isomerization in the gas phase of both FA and TFA, finding that, according to this mechanism, t-FA should be the only observable isomer \citep{delaconcepcion2021transcis}. At the same time, the situation is more complex in the case of TFA, where the isomerization is indeed possible in the interstellar medium. The reactivity of H with FA has been studied on the surface of dust grains \citep{Bisschop2007, Chaabouni2020} and in rare-gas matrix experiments \citep{Cao2014}. In the latter case, addition reactions are dominant. On grains, where several processes are competitive (\ce{H2} recombination, \ce{H} diffusion), seemingly contradictory results were found. One indicates that FA must be relatively inert to hydrogenation \citep{Bisschop2007} while the other supports the destruction of FA to form, e.g., \ce{CO2} and \ce{H2O}. In this context, \ce{H} abstraction reactions have not been considered and may happen in the event of an overall low reactivity. The \ce{H} abstraction reactions from t-FA were studied by some of us  \citep{Markmeyer2019} in the context of the HOCO radical formation, finding an overall slow reaction. No information on c-FA is available. The reactivity of TFA with H is much less studied, although intramolecular isomerization has been studied \citep{Kaur2014, delaconcepcion2021transcis}. In addition to isomerization, exchange reactions also favor deuterium enrichment \citep{Aikawa2012}. The deuterium fractionation of molecules is a good estimator of the chemical maturity of interstellar objects \citep{Bacmann2003,Aikawa2012, Ceccarelli:2014} and dedicated data is required to fine-tune our understanding of isotopic fractionation processes. We have also contributed to that goal in the present paper, extending our study to HD abstraction reactions.

We employed theoretical methods with an accurate treatment of quantum tunneling \citep{Meisner2017, Lamberts2017, Oba2018, Alvarez-Barcia2018, Molpeceres2021, Miksch2021, Molpeceres2021d} to investigate \ce{H} abstraction reactions from FA and TFA, both by H and D atoms. Moreover, the effect of the binding of the acids with the surface in the context of their chemistry was elucidated. Supporting experiments were used to constrain the viability of the slow hydrogenation and deuteration of t-FA and observations of deuterated isotopologues of FA are also included to expand our conclusions on some of the reactions that we present. After a description of the theoretical methods used in our work, we study the chemistry of the acids, providing activation energies, reaction energies, binding energies, and reaction rate constants for all the adsorbates. All the reaction rate constants are fitted to commonly used analytic expressions to ease the implementation into astrochemical models. In addition, we provide kinetic isotopic effects (KIE) for \ce{H} abstraction reactions promoted by D atoms.



\section{Methodology}


\subsection{Computational Details}

\subsubsection{Reaction rate constants} \label{sec:rate_constants_method}

We probed the reactivity of both isomers of FA and TFA by performing two series of independent calculations. In the first place, we studied the kinetics of the following \ce{H} abstraction reactions employing an implicit surface approach \citep{Meisner2017}:

\begin{align}
    \ce{c-FA + H &-> t-HOCO + H2}, \label{eq:1}\\
    \ce{t-FA + H &-> c-HOCO + H2}, \label{eq:2}\\
    \ce{c-TFA + H &-> OCHS + H2},  \label{eq:3}\\
    \ce{t-TFA + H &-> OCHS + H2},   \label{eq:4}\\
    \ce{c-TFA + H &-> t-OCSH + H2}, \label{eq:5}\\
    \ce{t-TFA + H &-> c-OCSH + H2}. \label{eq:6}
\end{align}

\begin{figure}
    \centering
        \includegraphics[width=\linewidth]{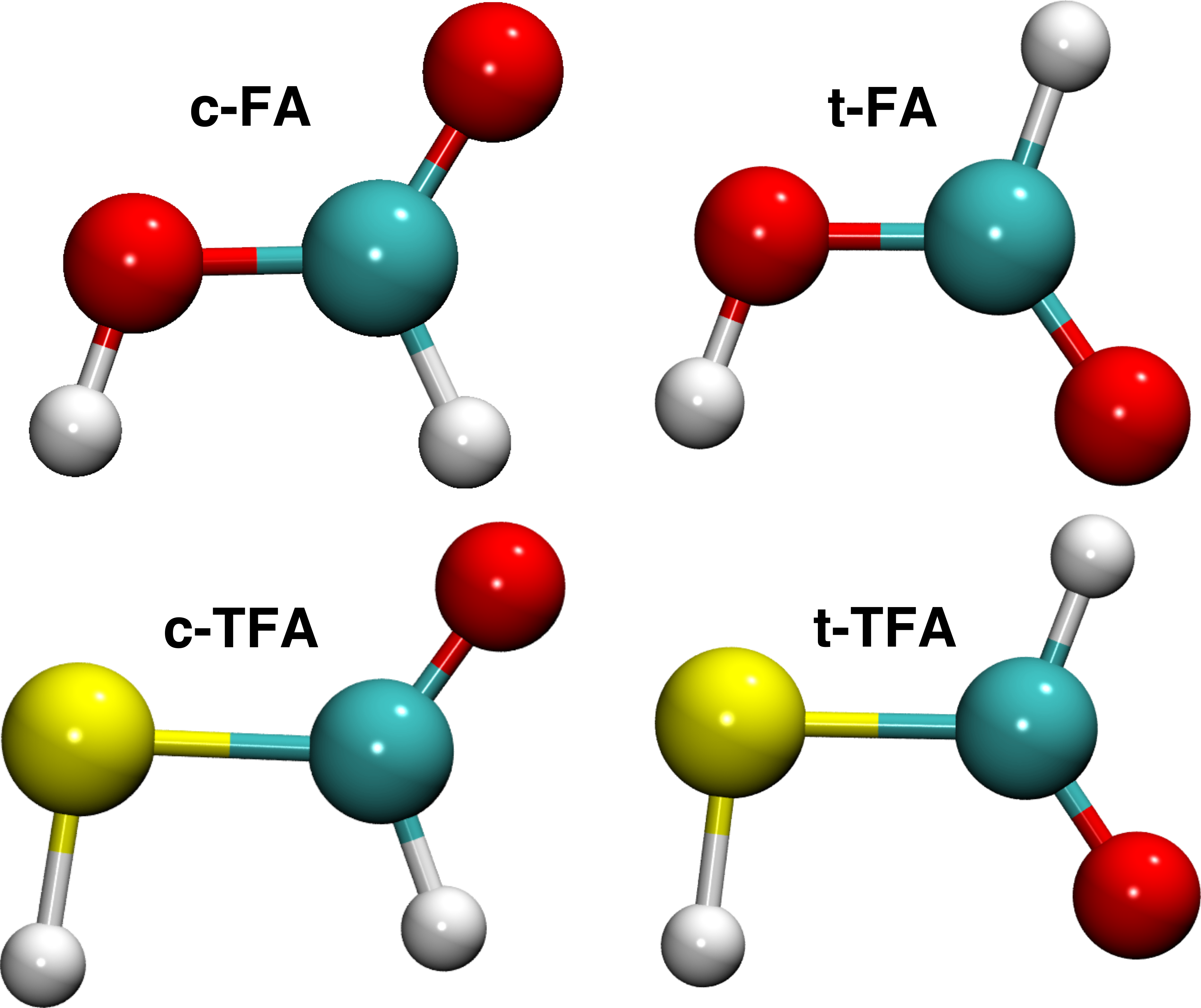} \\
    \caption{Molecular structure of the FA and TFA isomers.}
    \label{fig:structures}
\end{figure}

In the above reactions, c- and t- preceding the reactants indicate the cis (c) and trans (t) isomers of the reactant. In Figure \ref{fig:structures} we portray the structures of all reactants considered in this work. For TFA, we considered abstraction reactions both in the S-H and in the C-H moiety. Abstractions at the OH moiety are discarded on the basis of endothermicity (see below). To isomerize, the radicals formed on the surface can undergo subsequent hydrogenation:

\begin{align}
    \ce{HOCO + H &-> FA}, \label{eq:7}\\
    \ce{OCHS + H &-> TFA},  \label{eq:8}\\
    \ce{OCSH + H &-> TFA},   \label{eq:9}
\end{align}

where the isomer form of the products is deliberately left ambiguous. In this second step the parent molecule can be formed again as different isomer, depending on the orientation of the radical with the surface \citep{Molpeceres2021d}. 

For FA, preliminary explorations showed that the \ce{c,t-FA + H -> t,c-OCHO} reaction, e.g. abstraction of a H atom from the -OH moiety leading to the formation of the formate radical is endothermic irrespective of the parent FA isomer under consideration. For all the reactions (see Section \ref{sec:results}), we also investigated the HD abstraction reactions. The protocol that we followed to study the reactions is as follows. First, we optimized the classical transition states from reasonable guesses of potential energy surface (PES) obtained from relaxed scan calculations. Second, we run intrinsic reaction coordinate (IRC) calculations starting from the transition state structure to obtain a guess for each reaction's unimolecular reactant complex. The endpoints toward reactants of the IRC are later further optimized and used as reactant states for the rate constant calculations. Rate constants are then computed by means of semi-classical instanton theory \citep{lan67,mil75,col77,Rommel2011-2, Rommel2011} in the range from $T_{c}$ to 50 K, with $T_{c}$ being the crossover temperature for each reaction defined as the temperature where tunneling starts to dominate the rate constant value (see the formulation of \cite{Gillan1987}):

\begin{equation}
    T_\text{c} = \frac{\hbar \omega_{i}  }{2\pi k_\text{B}} .
\end{equation}

For reaction \eqref{eq:2}, we must stop at 65 K due to the difficulties of converging the instanton path below that temperature, as also found in a recent work \citep{Markmeyer2019}. These rate constants include a multidimensional treatment of quantum tunneling, which is a necessary trait for \ce{H} abstraction reactions. The effect of a surface was simulated with the implicit surface approach \citep{Meisner2017}, as mentioned above, which allows for affordable treatment of surface effects. For computing the instanton rate constants, we optimized an initial instanton path close to the crossover temperature of each reaction. From that, we computed the rate constant at lower temperatures until the final temperature of 50~K (or 65~K in the case of t-FA, see above). There are no differences in the protocol between reactions involving \ce{H2} and \ce{HD}. For all electronic structure calculations, we used the Gaussian16 \citep{g16} suite of programs interfaced with DL-Find/ChemShell \citep{Quasi, Kastner2009,Chemshell}. The level of theory employed throughout the rate constant calculations is MPWB1K/def2-TZVP \citep{Zhao2005, Weigend2005}, which rendered activation energies in excellent agreement (\textless 1 kcal mol$^{-1}$) with more expensive UCCSD(T)-F12a/cc-pVTZ calculations that we considered as our reference calculations. This level of theory was also ranked among the best ones in a previous study \cite{Markmeyer2019}. This functional behaves slightly worse for thermochemistry and the prediction of reaction energies, but such a magnitude does not interfere with the values of the rate constants. We nonetheless report also CCSD(T)-F12a/cc-pVTZ for this magnitude.

\subsubsection{Binding Site Estimation} \label{sec:binding_method}

We estimated the binding orientations for both isomers of FA and TFA from computational simulations explicitly accounting for the interaction between adsorbate and surface, in contrast to section \ref{sec:rate_constants_method}. The adsorbate orientation on a surface is fundamental to understanding reactivity. Some directions may promote or inhibit the reactions in specific places within the molecule. As a matter of fact, we recently found orientation effects during the formation of TFA from OCS due to the formation of hydrogen bonds of the OCSH radical with the surface \citep{Molpeceres2021d}. 

To determine the binding sites of the different adsorbates, we followed a sequential scheme following our previous works \citep{Molpeceres2021}. First, from a randomly generated water cluster containing 46 \ce{H2O} molecules, we performed one molecular dynamics heating cycle at 300~K for 50 ps to amorphize the structure. From this simulation, we extracted 5 frames (one every 10~ps) and cooled them down to 10~K each for 10 ps. Each one of these structures is then further relaxed at the GFN2-xTB level of theory \citep{Grimme2017, Bannwarth2019}. This method allows us for fast and reliable exploration of adsorption geometries of closed-shell molecules, such as the ones of this study \citep{Germain2021}. In addition, we refined the electronic energies of the optimized structures with density functional theory (DFT) energies coming from the PW6B95-D3BJ/def2-TZVP method \citep{Zhao2005, Weigend2005, Grimme2010, Grimme2011}. The choice of this functional is motivated by the improved thermochemistry associated with it \citep{Zhao2005}. With the structural models for the ice ready, we placed the molecules of interest, c/t-FA, and c/t-TFA, on the cluster, with the molecules' centers of mass on equispaced points of an ellipsoid spanning a Fibonacci lattice. The initial distance between the molecules' center of mass and the cluster is approximately 3~\AA. We placed 12 adsorbates per structural model following this procedure, making a total of 12x5(structural models)x4(adsorbates)=240 initial structures. We finally relaxed the adsorbate+cluster structure for all of them. To ease the convergence of the structures, we fixed the atoms that were not at a minimum distance of 9~\AA~to any adsorbate atom at the beginning of the calculation. With all the necessary ingredients, binding energies $\Delta E_{\text{bin}}$ can be computed as:

\begin{equation}
    \Delta E_{\text{bin}} = E_{\text{Ice+Ads}} - (E_{\text{Ice}} + E_{\text{Ads}}).
\end{equation}

In the above equation, $E_{\text{Ice+Ads}}$ corresponds to the potential energy of the adsorbate and ice complex at short distance, $E_{\text{Ice}}$ to the potential energy of the water cluster, and $E_{\text{Ads}}$ to the potential energy of the adsorbate under consideration. Note that in a real ice, more variety of adsorption sites would be possible due to the codeposition of other molecules, e.g. \ce{CO} on the ice. In this work, we are interested in the reactivity of FA and TFA with H/D and the possible orientation effects that stem from the interaction of the adsorbates with the surface leaving a particular molecular face available or forbidden for reaction. We have analyzed this effect in two steps. First, from the histogram of binding energies we took the structures corresponding to the high binding situations (e.g, higher than the average, $\Delta E_{\text{bin}} > \Delta \overline{E}_{\text{bin}}$), that are most likely to be populated under ISM conditions. Second, we measured the distance of any of the atoms susceptible to forming a hydrogen bond (all hydrogens and the carboxylic oxygen) with the closest water molecule for each one of these configurations. A value lower than $\sim$2.2~\AA~ indicates the formation of a localized hydrogen bond, locking that particular bond, and thus the orientation for reaction. The formation of a hydrogen bond precludes the formation of an activated complex, due to the additional required energy to break the hydrogen bond between surface and adsorbate. The criterion that we apply is that each immobilized H atom from the adsorbate cannot react. 

\subsection{Experimental Setup}

As shown in section \ref{sec:rate_constants}, for reaction \ref{eq:2} we found a low rate constant of around 1\e{-1} s$^{-1}$, as well as very similar behavior for \ce{HD} abstraction. To experimentally check the viability of the reaction, experiments were performed using an Apparatus for Surface Reactions in Astrophysics (ASURA) at Hokkaido University. The ASURA mainly consists of a stainless-steel reaction chamber, an atomic source chamber, a quadrupole mass spectrometer, a Fourier-transform infrared spectrometer (FTIR), and an Al reaction substrate courted with amorphous silicates (\ce{Mg2SiO4}) layers. More details can be found in \cite{Nguyen2020}. The substrate temperature can be controlled between 5 to 300 K. Gaseous FA and D atoms were continuously co-deposited onto the pre-deposited porous amorphous solid water with the thickness of $\sim$20 monolayers (MLs: 1 ML = 1\e{15} molecules cm$^{-2}$) at 10~K. The deposition rate of FA was 7.8\e{11} molecules cm$^{-2}$ s$^{-1}$ and the D-atom flux was 4.6\e{14} atoms cm$^{-2}$ s$^{-1}$, where the FA/D ratio was 1.7\e{-3}. D atoms were produced through the dissociation of \ce{D2} in a microwave-induced plasma in the atomic chamber and cooled to 100~K before landing on the substrate. Reaction products were monitored in-situ by the FTIR.

\subsection{Observations}

We have used the high-sensitivity observations carried out by \citet{JimenezSerra2016} toward the position of the dust peak in the L1544 pre-stellar core. This core is on the verge of gravitational collapse and it has experienced a strong depletion of CO, which induces deuterium fractionation as measured from deuterated species such as \ce{N2D+}. Therefore, we have selected the position of the dust peak of L1544 because of the large D/H ratio measured toward this position \citep[of $\sim$25\%; see][]{redaelli19}. The observations toward L1544 were carried out in the 3 mm  band using the Instituto de Radioastronom\'{\i}a Milim\'etrica (IRAM) 30 m telescope. The achieved rms noise level was 1.6-2.8 mK for the dust peak position in L1544. The details of these observations can be found in \citet{JimenezSerra2016}. Note, however, that they did not cover the full 3 mm band. 

In order to cover the frequency gaps of the high-sensitivity 3mm survey of \citet{JimenezSerra2016}, we also used the L1544 spectroscopic survey obtained within the ASAI (Astrochemical Surveys At IRAM) Large Program \citep[see][]{lefloch17}. The typical rms noise level achieved in the 3mm band survey was 4-7 mK. The details of the observations can be found in \citet[][]{lefloch17}.


\section{Results} \label{sec:results}

\subsection{\ce{H} Abstraction Rate Constants} \label{sec:rate_constants}

In Table \ref{tab:summary} we present all the energetic parameters of reactions \ref{eq:1} to \ref{eq:6}. A detailed description for each reaction follows in the next subsections.

\begin{table}
\caption{Reaction energies (including ZPE, $\Delta U_{r}$ in kJ mol$^{-1}$) at the MPWB1K/def2-TZVP level and at UCCSD(T)-F12/cc-pVTZ-F12 for the electronic energies (in parenthesis), activation energies (including ZPE, $\Delta U_{a}$) at the same level  and crossover temperatures (T$_{c}$ in K) for reactions \ref{eq:1} to \ref{eq:6} }             
\label{tab:summary} 
\centering                          

\begin{tabular}{c c c c c }        
\hline\hline                 
Reaction & $\Delta U_{r}$ & $\Delta U_{a}$ & T$_{c}$ \\    
\hline                        
   Reaction \ref{eq:1} & -37.8 (-44.1) & 27.0 (30.3) & 362 \\
   Reaction \ref{eq:2} & -19.2 (-27.3) & 40.2 (43.2) & 367 \\
   Reaction \ref{eq:3} & -57.8 (-63.9) & 10.6 (9.9)  & 233 \\
   Reaction \ref{eq:4} & -54.5 (-61.3) & 10.3 (9.8)  & 236 \\
   Reaction \ref{eq:5} & -72.2 (-77.9) & 12.9 (14.8) & 297 \\
   Reaction \ref{eq:6} & -60.1 (-66.7) & 18.6 (20.3) & 350 \\
    
\hline                                   
\end{tabular}
\end{table}

\subsubsection{Formic Acid (Theoretical Results)}

\begin{figure}
    \centering
        \resizebox{\linewidth}{!}{\includegraphics[width=1\linewidth]{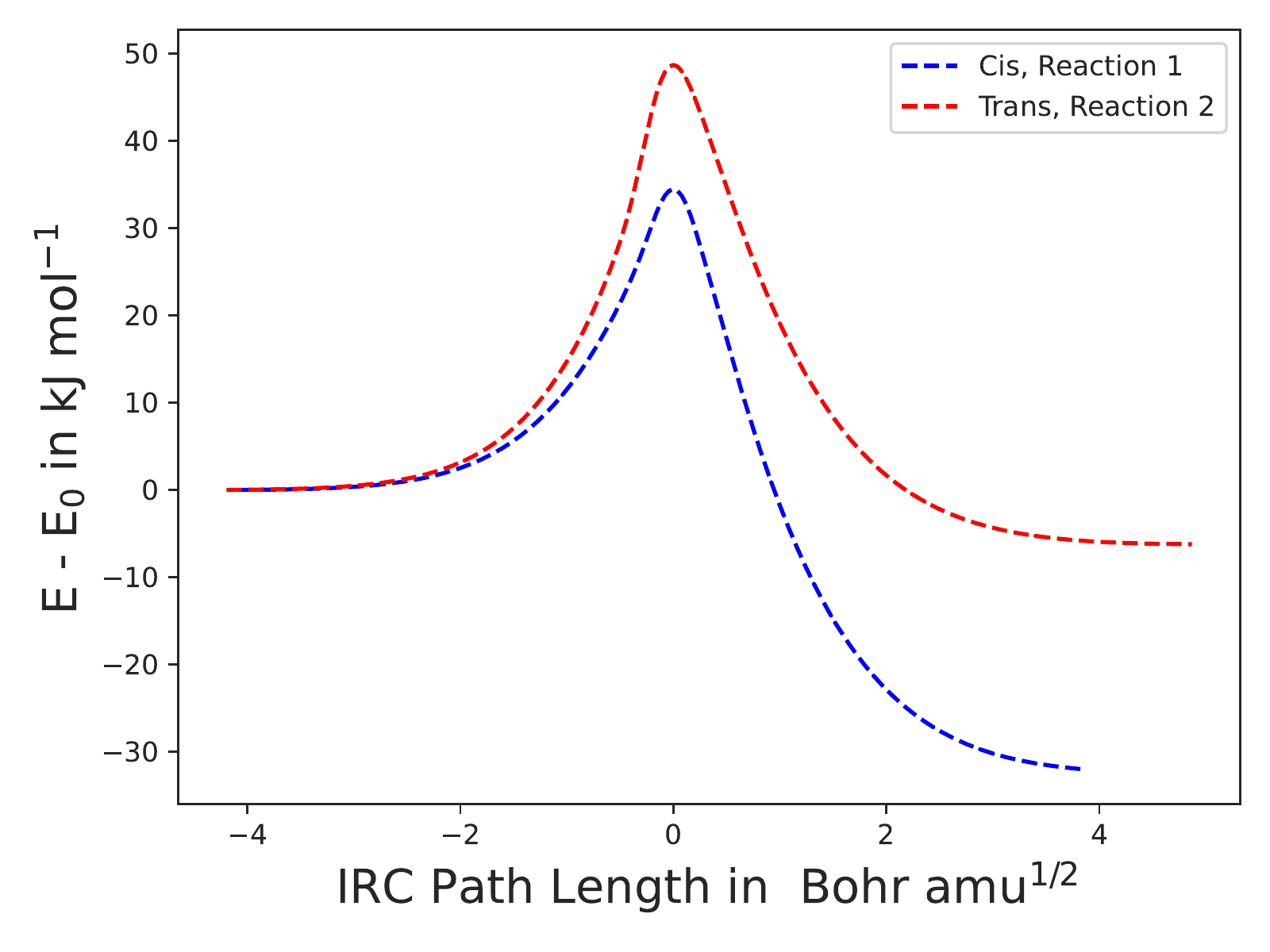}} \\
    \caption{Intrinsic reaction coordinate calculations for reactions \ref{eq:1} and \ref{eq:2} as described in the main text. Displayed energies are not corrected with zero point energy contributions.}
    \label{fig:FAIRC}
\end{figure}

\begin{figure}
    \centering
        \includegraphics[width=\linewidth]{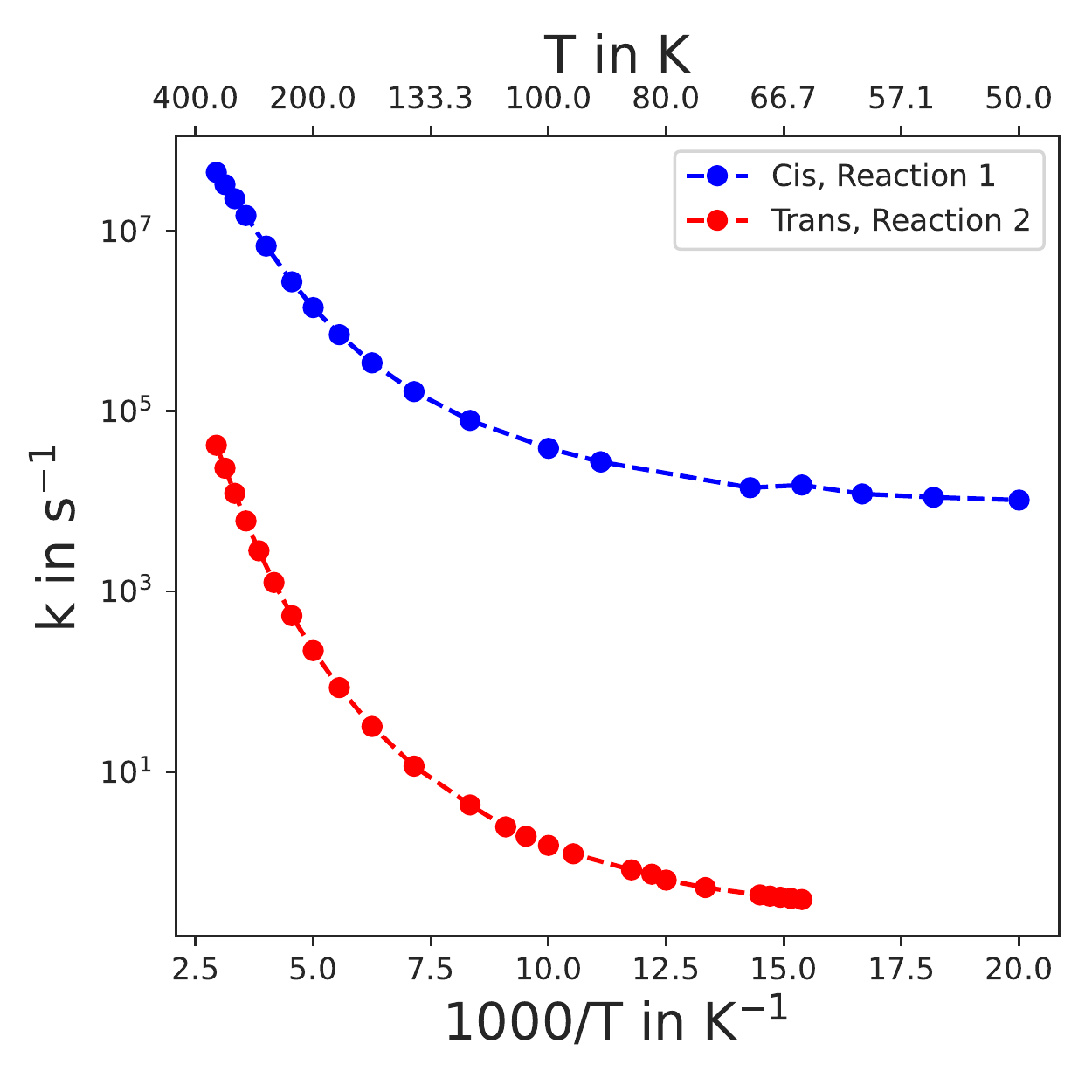} \\
    \caption{Instanton reaction rate constants below $T_\text{c}$~for reactions \ref{eq:1} and \ref{eq:2}. }
    \label{fig:KFA}
\end{figure}

The activation energies including ZPE contributions ($\Delta U_{a}$) for reactions \ref{eq:1} and \ref{eq:2} are 27.0 kJ mol$^{-1}$ (30.3 kJ mol$^{-1}$ at the UCCSD(T)-F12/cc-pVTZ-F12 level) and 40.2 (43.2) kJ mol$^{-1}$. The activation energies in parenthesis refer always to UCCSD(T)-F12/cc-pVTZ-F12. Both activation energies are sufficiently separated to justify a preferential reactivity on the c-FA, which will be further developed in light of the orientation effects presented in the next section. The energy along the intrinsic reaction coordinate of both reactions is presented in Figure \ref{fig:FAIRC}. From the plot, we also observe that the \ce{H} abstraction reaction is exothermic for both reactions with unimolecular reaction energies obtained from the optimization of the IRC endpoints of $-$19.2 (-27.3) kJ mol$^{-1}$, in contrast with the values of c-FA, $-$37.8 (-44.1) kJ mol$^{-1}$. This last result correlates with the value of the barrier and confirms preferential chemistry for the cis isomer. The different reactivity is, of course, also reflected in the different rate constants for the system, presented in Figure \ref{fig:KFA}. Our results agree with the values for the only available reaction (Reaction \ref{eq:2}) \citep{Markmeyer2019}. As mentioned during Section \ref{sec:rate_constants_method}, we could not converge instanton paths below 65~K for t-FA. From Figure \ref{fig:KFA} we observe, however, that the reaction rate constants reasonably converge to the tunneling asymptotic limit. 

The $\sim$13 kJ mol$^{-1}$ gap in the activation energies between the cis and trans isomer translates in more than 5 orders of magnitude difference in the reaction rate constants. This renders radically different kinetics for the reaction, favoring the destruction of c-FA over t-FA through \ce{H} abstraction reactions. The shape of the barrier depicted in Figure \ref{fig:FAIRC} also show that the barrier for reaction \ref{eq:2} is notably wider, which has a consequential effect in the rate constants as well.

\begin{figure}
    \centering
        \includegraphics[width=\columnwidth]{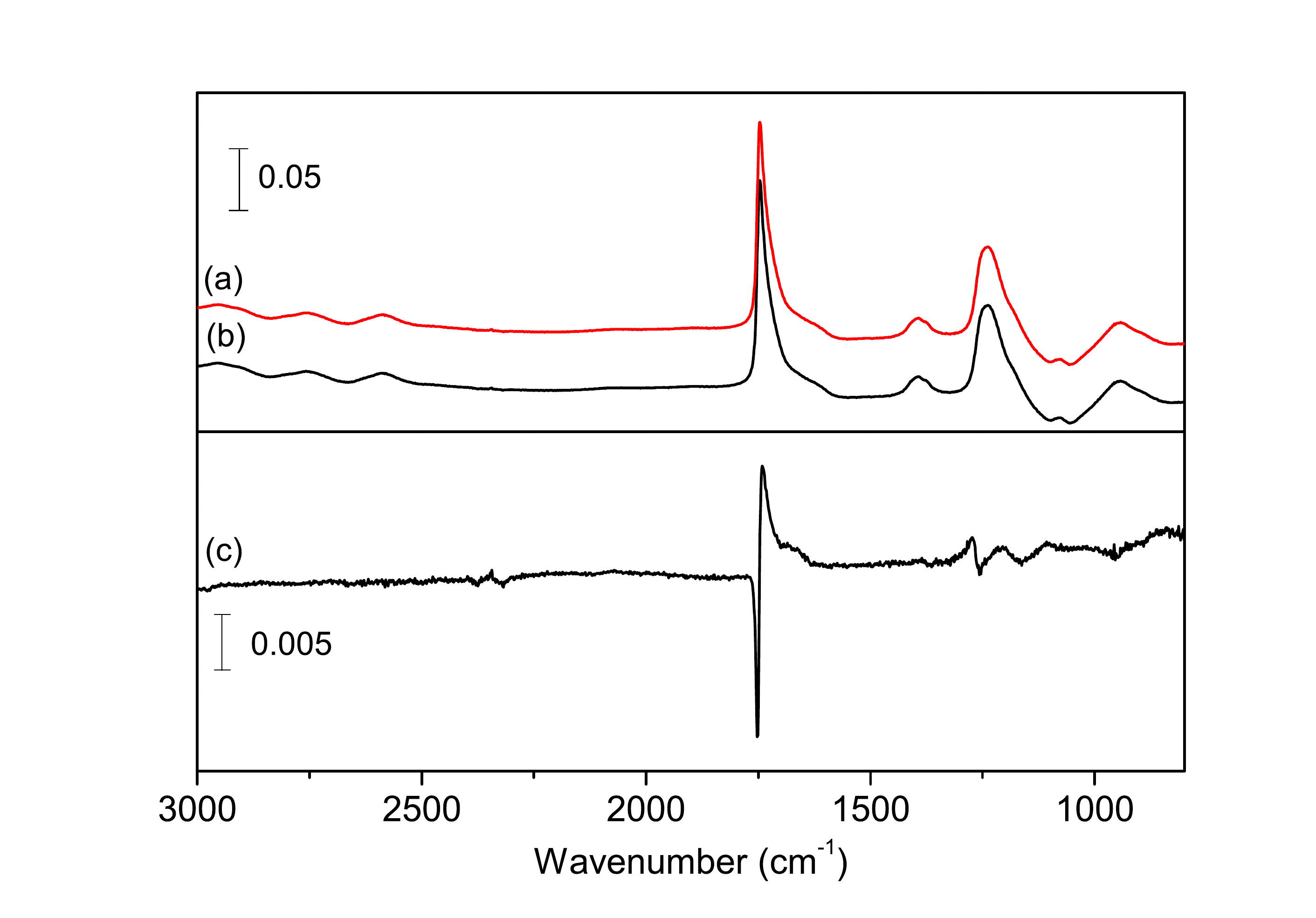} \\
    \caption{FTIR spectra obtained after (a) codeposition of FA and D atoms and (b) FA for 3 h at 10 K, and (c) the difference spectrum between spectra (a) and (b).}
    \label{fig:EXP}
\end{figure}

\subsubsection{Formic Acid (Experiments)}

Another question that may arise is whether reaction \ref{eq:2} really happens or if it is not competitive with H diffusion or addition reactions \citep{Bisschop2007, Chaabouni2020}. A rate coefficient of 10$^{-1}$ s$^{-1}$ for the reaction is below the hopping rates for H diffusion found for average binding sites on amorphous solid water as presented by \cite{Hama2012b}. Therefore, determining the viability of the reaction against diffusion is not trivial from static quantum chemical calculations. We have performed experiments with deuterium at low temperatures to check whether the reaction proceeds profiting from the small KIE of reaction \ref{eq:2} (see Section \ref{sec:deuterium}). The appearance of the C-D band at 2000--2200 cm$^{-1}$ (typical wavenumbers for C-D stretching band) will indicate that reaction \ref{eq:2} takes place, followed by the addition of a D atom to the formed HOCO radical yielding DCOOH. Considering that FA is deposited in the UHV chamber at room temperature, it is safe to think that all FA is t-FA, taking into account an equilibrium constant (trans/cis) of 10$^{3}$ at 300~K \citep{delaconcepcion2021transcis}. In section \ref{sec:discussion} we also present that H abstraction by deuterium in t-FA has very similar kinetics to \ce{H} abstraction by hydrogen atoms, and such a reaction will serve as an adequate proxy of reaction \ref{eq:2} in our experiments.

The results of the experiment, after 3h-codeposition of FA and D atoms, are presented in Figure \ref{fig:EXP}. We do not clearly observe a C-D stretching band at 2000-2200 cm$^{-1}$, and thus we consider reaction \ref{eq:2} non-viable under interstellar conditions. On the contrary, the abstraction of c-FA (equation \ref{eq:1}) can not be directly observed under laboratory conditions owing to the large trans/cis equilibrium constant of FA, but experiments using the same setup show that rate constants in the order of $k(50)\sim$10$^{4}$ s$^{-1}$ are observable, as in the case of the hydrogenation of OCS \citep{Molpeceres2021d, Nguyen2021}.

\subsubsection{Thioformic Acid}

In contrast with FA, hydrogen can be abstracted from TFA from two different positions of its structure, namely in the thiol group (SH) and in the  group (CH moiety). This is due to both hydrogen abstractions being exothermic, in contrast with the endothermic pathways found for the \ce{H} abstraction in the OH group of formic acid, an analogy of what happens in \ce{H} abstraction reactions on \ce{H2O} and \ce{H2S}.

\begin{figure}
    \centering
        \includegraphics[width=\linewidth]{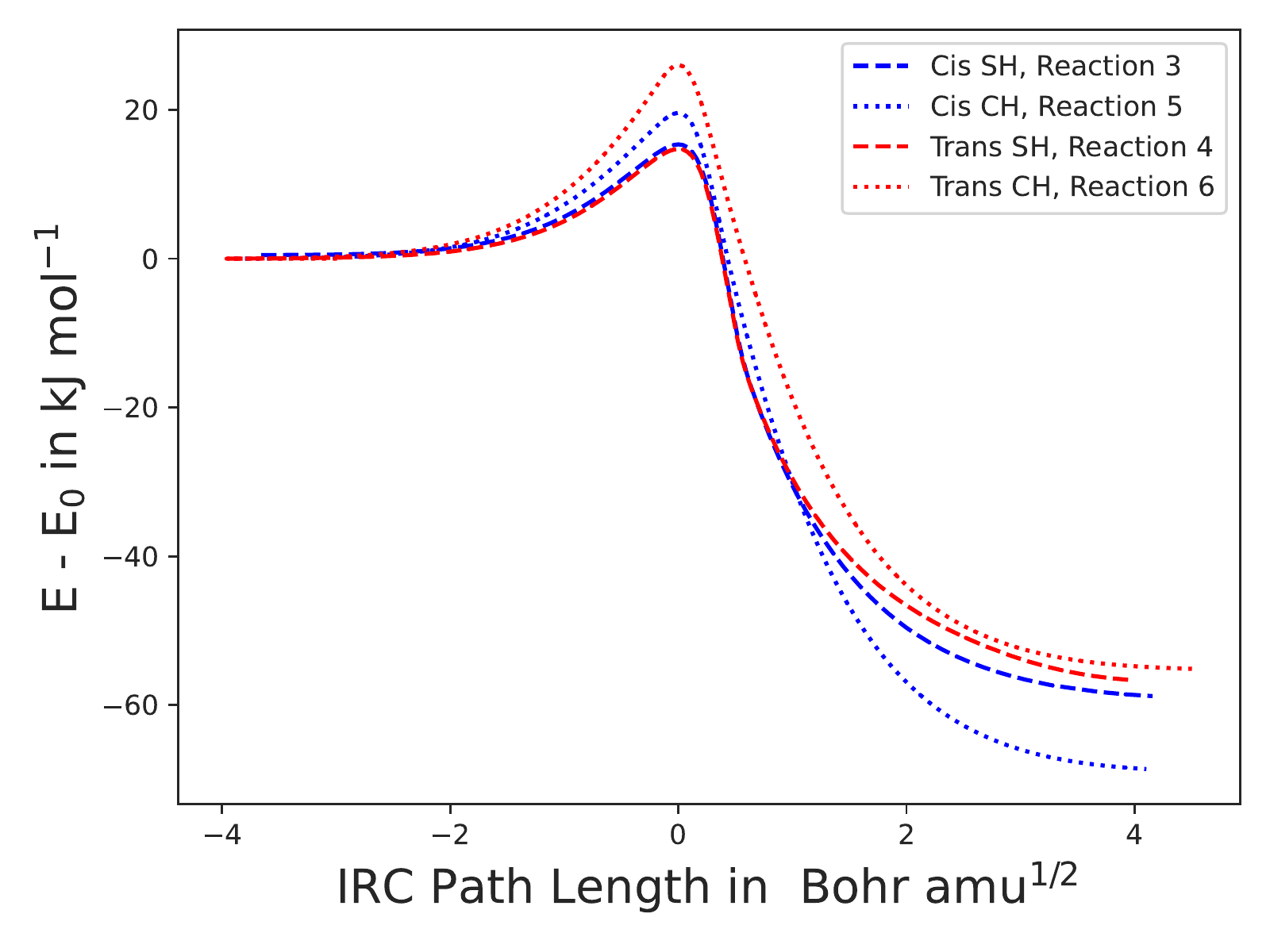} \\
    \caption{Intrinsic reaction coordinate calculations for reactions \ref{eq:3} to \ref{eq:6} as described in the main text. Displayed energies are not corrected with zero point energy contributions.}
    \label{fig:TFAIRC}
\end{figure}

In c-TFA, the \ce{H} abstraction in the thiol group presents an activation barrier of 10.6 (9.9) kJ mol$^{-1}$ and the abstraction on the thiol group of t-TFA has a barrier of 10.3 (9.8) kJ mol$^{-1}$, two very similar values. This is in contrast with the abstraction reactions on the CH moiety. Reaction \ref{eq:5} has an activation barrier of 12.9 (14.8) kJ mol$^{-1}$ whereas reaction \ref{eq:6} presents a higher activation energy of 18.6 (20.3) kJ mol$^{-1}$. The IRC for both abstractions is depicted in Figure \ref{fig:TFAIRC}. We observe an identical reaction profile for the abstractions in the SH moiety in c/t-TFA. This is also in agreement with the very similar reaction energies of -57.8 (-63.9) and -54.5 (-61.3) kJ mol$^{-1}$ for the cis and trans isomer, obtained from the IRC structures, as mentioned above. Both abstractions at the SH moiety present lower activation energies than their counterparts at the CH one. This trend is also coherent for the reaction energies, whose values are of -72.2 (-77.9)  kJ mol$^{-1}$ for c-TFA and -60.1 (-66.7) kJ mol$^{-1}$ for t-TFA using the same protocol, for the \ce{H} abstraction at the CH moiety.

\begin{figure}
    \centering
        \includegraphics[width=\linewidth]{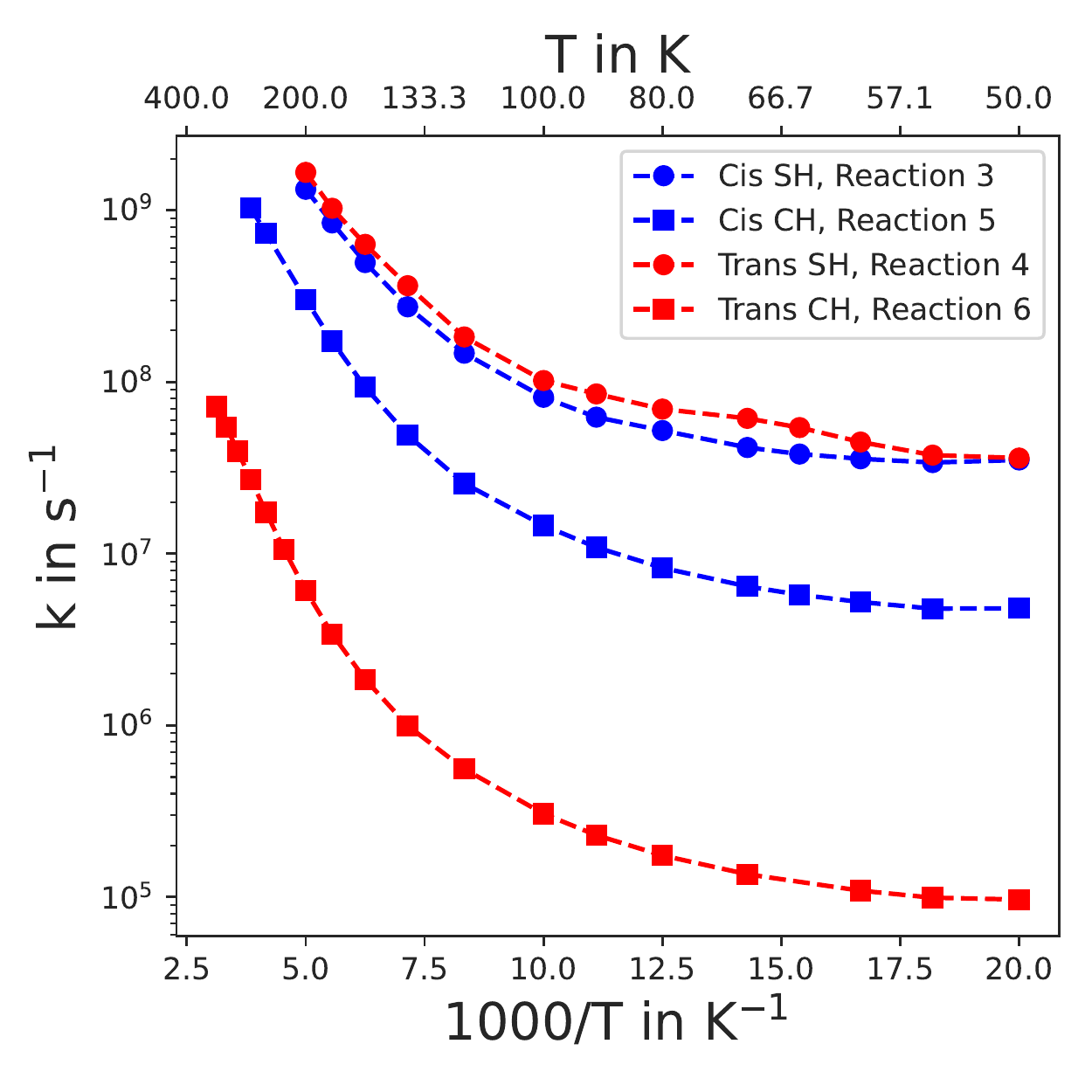} \\
    \caption{Instanton reaction rate constants below $T_\text{c}$~for reactions \ref{eq:3} to \ref{eq:6}. }
    \label{fig:KTFA}
\end{figure}

We portray the reaction rate constants for each reaction in Figure \ref{fig:KTFA}. The results confirm the trend that we found for the activation energies. Although the abstraction in the thiol group does not show any particular trend, there is preferential destruction of c-TFA through reaction \ref{eq:5} to reaction \ref{eq:6}. The specific importance of this reaction in the enrichment of t-TFA on dust requires information on the particular binding of c/t-TFA with the surface that we cover in the next section. 

All the reactions on TFA present reaction timescales that should be observable in experiments. However, we have been unable to acquire pure TFA to conduct direct hydrogenation/deuteration experiments, and the conclusion must be made based on our theoretical results.

\subsubsection{Fit to analytic expression}

To ease the incorporation of the rate constants presented here into astrochemical models, we have fitted the results of the previous section to a modified Arrhenius--Kooij equation suited to account for the asymptotic profile of rate constants including quantum tunneling \citep{Zheng2010}:

\begin{equation}
    k = \alpha\left(\dfrac{T+T_{0}}{300~\text{K}}\right)^{\beta}\exp\left(-\gamma\dfrac{T+T_{0}}{T^{2} + T_{0}^2}\right).
\end{equation}

We fitted the parameters: $\alpha$ (s$^{-1}$), $T_{0}$ (K), $\beta$ and $\gamma$ (K) for reactions \ref{eq:1} to \ref{eq:6} and the results are gathered in Table \ref{tab:FITH}. Our fits deteriorate at low temperatures when a lot of points are included toward high temperatues, due to an over-representation of these points. Therefore and since we are interested in rate constants at low temperatures, we restricted the values of the fit to the rate constants below 250~K.

\begin{table}
\caption{Fitted parameters of the modified Arrhenius-Kooij equation for reactions \ref{eq:1} to \ref{eq:6}}             
\label{tab:FITH} 
\centering                          
\begin{tabular}{c c c c c}        
\hline\hline                 
Reaction & $\alpha$ (s$^{-1}$) & $\beta$ & $\gamma$ (K) & $T_{0}$ (K) \\    
\hline                        
   Reaction \ref{eq:1} & 1.21\e{8} & 5.76 & 1.03\e{3} & 1.65\e{2}\\
   Reaction \ref{eq:2} & 1.82\e{3} & 11.77 & 8.67\e{2} & 1.72\e{2} \\
   Reaction \ref{eq:3} & 9.49\e{11} & -2.09 & 1.09\e{3} & 1.16\e{2} \\
   Reaction \ref{eq:4} & 1.09\e{11} & 1.49 & 7.36\e{2} & 1.25\e{2} \\
   Reaction \ref{eq:5} & 7.51\e{11} & -2.09 & 1.31\e{3} & 1.23\e{2} \\
   Reaction \ref{eq:6} & 8.18\e{8}  & 3.22 & 9.69\e{2} & 1.52\e{2} \\

\hline                                   
\end{tabular}
\end{table}

\subsection{Binding Energy Estimation: Orientation Effects} \label{sec:binding_energies}

\subsubsection{Formic Acid}

\begin{figure}
    \centering
        \includegraphics[width=\linewidth]{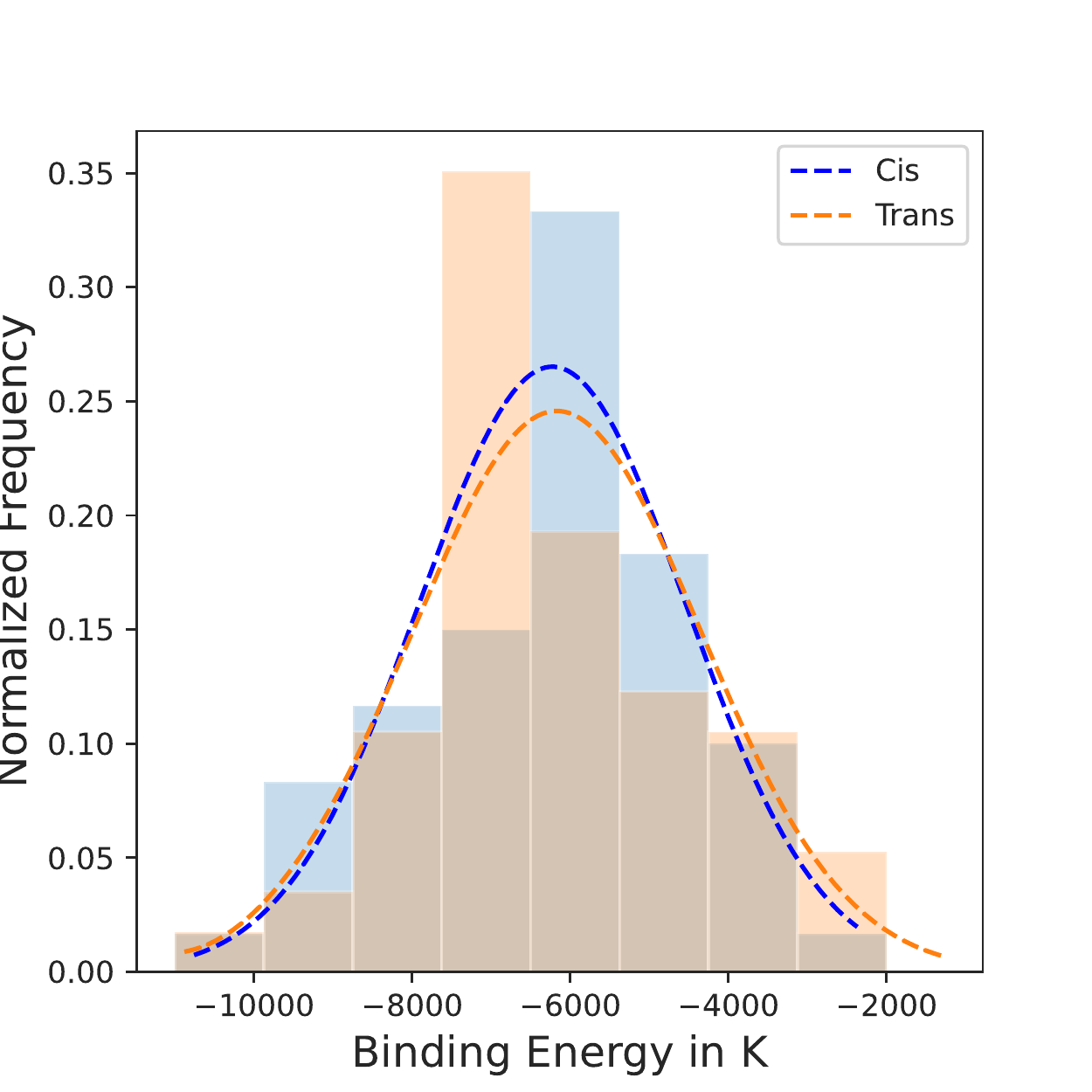} \\
    \caption{Binding energies of cis/trans-Formic acid. Dashed line - Gaussian fit of the binding energies. $\mu_{\text{cis}}$=-6227~K, $\sigma_{\text{cis}}$=1692~K. $\mu_{\text{trans}}$=-6170~K, $\sigma_{\text{trans}}$=1826~K.}
    \label{fig:FAbin}
\end{figure}

\begin{figure}
    \centering
        \includegraphics[width=\linewidth]{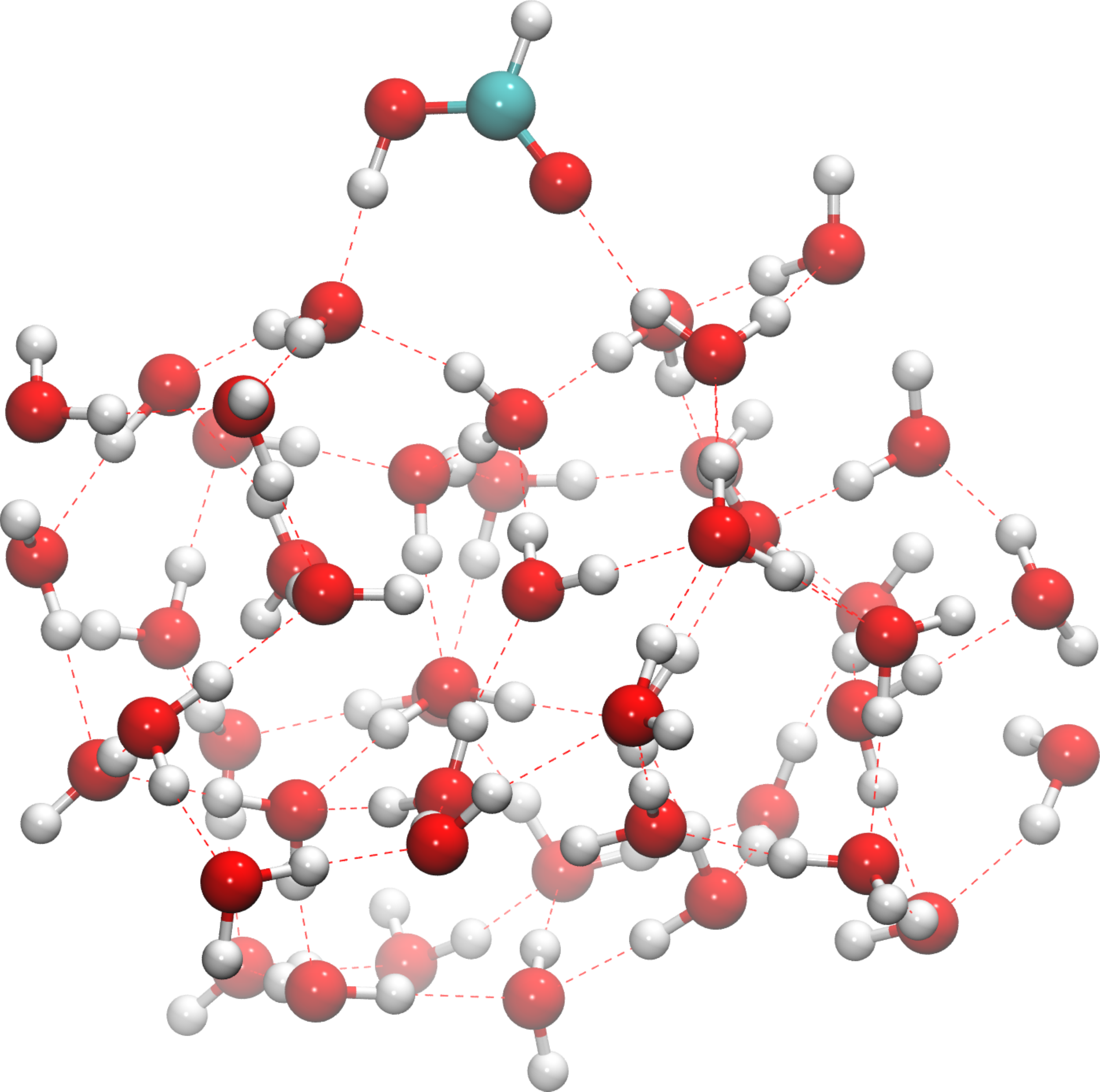} \\
    \caption{Examplary binding orientation of t-FA on ASW. In the figure, we depict the clear localization in the binding orientation of the molecule forming two hydrogen bonds with different atoms of the surface.}
    \label{fig:FAExam}
\end{figure}

In addition to the reaction rate constants, orientation effects also drive the reactivity on ASW, especially considering that both FA and TFA are polar molecules. We determined the most favorable binding sites based on a distance criterion after computing the binding energies as described in Section \ref{sec:binding_method}. For FA, we present these binding energies in Figure \ref{fig:FAbin} with average binding energies $\mu$ and standard deviation $\sigma$. We observe hardly any difference in the binding energy distribution for both isomers. The average value for the binding energy, of around -6200~K, and the distributions are in very much concordance with the values already reported in the literature for pure DFT computations \citep{Ferrero2020}. A closer analysis on the possibility of hydrogen bonding revealing orientation effects is summarized in Table \ref{tab:locFA}. The results show that the binding in formic acid is considerably localized, with the hydroxyl group permanently binding with the surface. Furthermore, in the case of t-FA, the C=O moiety is also always forming a bond with the surface, see Figure \ref{fig:FAExam}.

\begin{table}
\caption{Fraction of different localized binding for the high binding energy adsorption sites in the c/t-FA/\ce{H2O} system. Values in parentheses represent the intervals of confidence of our estimation employing a binomial distribution and 95\% confidence interval.}             
\label{tab:locFA} 
\resizebox{\linewidth}{!}{
\centering                          
\begin{tabular}{c c c c}        
\hline\hline                 
System & HCO--HOH & COH--OHH & OCH--OHH \\    
\hline                        
   c-FA & 0.59 (0.38--0.77) & 1.00 (0.89--1.00) & 0.00 (0--0.11) \\
   t-FA & 1.00 (0.93--1.00) & 1.00 (0.93--1.00) & 0.00 (0--0.11) \\
\hline                                   
\end{tabular}}
\end{table}

\subsubsection{Thioformic Acid}

\begin{figure}
    \centering
        \includegraphics[width=\linewidth]{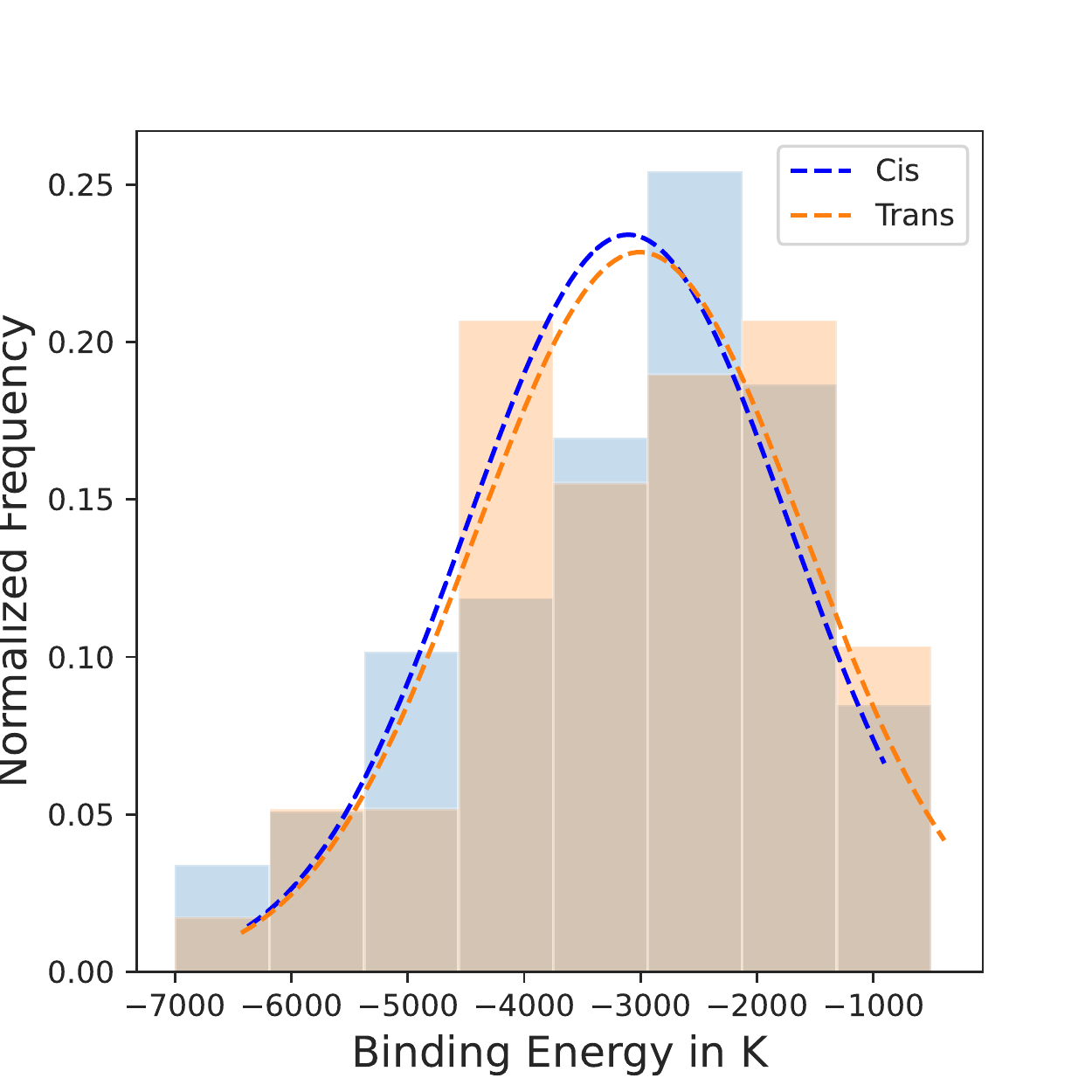} \\
    \caption{Binding energies of cis/trans-Thioformic acid. Dashed line - Gaussian fit of the binding energies. $\mu_{\text{cis}}$=-3105~K, $\sigma_{\text{cis}}$=1384~K. $\mu_{\text{trans}}$=-3005~K, $\sigma_{\text{trans}}$=1418~K. }
    \label{fig:TFAbin}
\end{figure}

The distribution of binding energies for the different isomers of TFA is presented in Figure \ref{fig:TFAbin}. The binding energies are, with average values of $-3105$ K for c-TFA and $-3005$ K for t-TFA, lower than in the case of FA and are very similar to the values of the OCSH radical \citep{Molpeceres2021d}. Again the difference of binding energies between isomers is within the standard deviation values, and there is no significant change in the affinity of the molecules with the surface. As in the case of FA, we have estimated the preferential binding mode for the binding sites with higher-than-average binding energies. The results are summarized in Table \ref{tab:locTFA}. Contrary to the situation found for FA, for TFA, not all the high binding situations are dominated by the interaction of the SH moiety with a water molecule, and some configurations leave room for reaction in the SH moiety. We found a similar behavior for the binding situations through the CO moiety. Another difference regarding FA appears when we determine the possibility of bonding through the CH group, which leads to a small, yet noticeable amount of binding situations ($\sim$7\%). Finally, we found that c-TFA forms a hydrogen bond with ASW through the -SH moeity in a 88\% of the high binding situations, in contrast with t-TFA where this quantity goes down to 63\%, indicating that reactions in the thiol group are a 25\% more likely in t-TFA (see Table \ref{tab:locTFA}, column 3). 

A closer inspection of the structures reveals that the automated criterion is valid for a qualitative estimation of orientation effects. 

\begin{table}
\caption{Fraction of different localized binding for the high binding energy adsorption sites in the c/t-TFA/\ce{H2O} system. Values in parentheses represent the intervals of confidence of our estimation employing a binomial distribution and 95\% confidence interval.}             
\label{tab:locTFA}      
\resizebox{\linewidth}{!}{
\centering                          
\begin{tabular}{c c c c}        
\hline\hline                 
System & HCO--HOH & CSH--OHH & OCH--OHH \\    
\hline                        
   c-TFA & 0.77 (0.58--0.90) & 0.88 (0.72--0.97) & 0.07 (0.02--0.22) \\
   t-TFA & 0.85 (0.69--0.95) & 0.63 (0.44--0.79) & 0.07 (0.02--0.22) \\
\hline                                   
\end{tabular}}
\end{table}

\


\section{Discussion} \label{sec:discussion}

\subsection{Hydrogen Abstraction of FA and TFA}

\begin{figure}
    \centering
        \includegraphics[width=\linewidth]{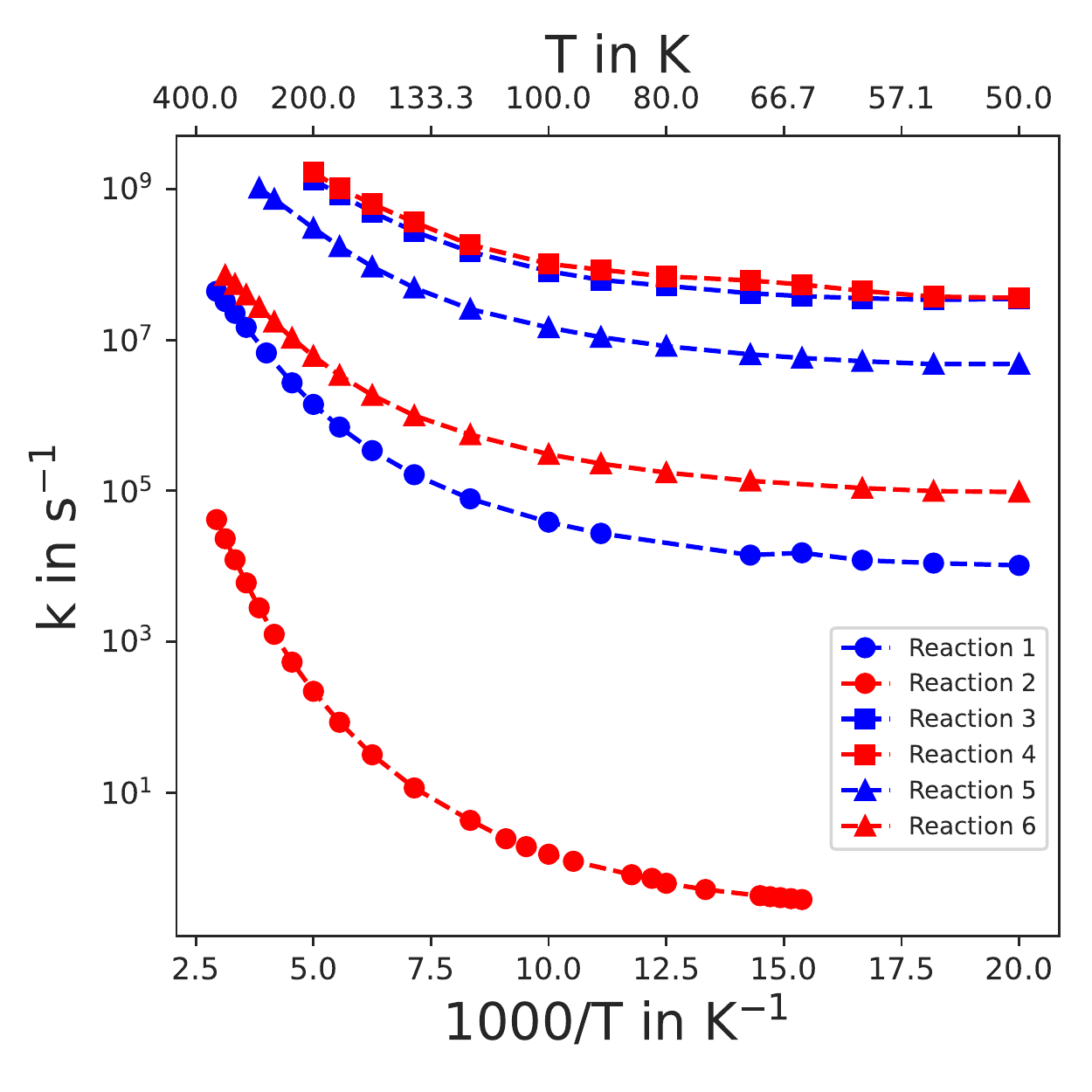} \\
    \caption{Rate constants for all the \ce{H} abstraction reactions considered in the present work.}
    \label{fig:HRates}
\end{figure}

The susceptibility of both FA and TFA to \ce{H} abstraction reactions was found with kinetic (rate constants) and thermodynamic (binding sites) arguments. As explained during Section \ref{sec:rate_constants_method}, the values of Table \ref{tab:locFA} and \ref{tab:locTFA} refer to the fraction of orientations that prevent or enable a particular reactivity. In the case of reaction in the CH moiety, the attack site is available when no CH bonds are partaking in weak bindings with the surface. Similarly, when the SH moiety is involved, such an orientation is precluded when the SH bond is directly interacting with the surface. To ease the discussion, we present all the rate constants that we have presented in section \ref{sec:rate_constants} in Figure \ref{fig:HRates}. 

From the rate constants, we can extract several points for the isomerization of acids on the surface of interstellar grains in interstellar clouds. We have confirmed a preferential reactivity based on isomerism by studying their interaction with hydrogen atoms. This preference emanates from two different factors, kinetic effects, and orientation effects. The short-range activation barriers for the reactions are different between isomers, a trend that was already observed for some reactions in a previous study \citep{Shingledecker2020}. This factor is usually the most important one, given the exponential dependence of the reaction rate constants on the magnitude of the activation energies and such is the case of \ce{H} abstractions occurring on FA. However, we observe that the preferential binding has a significant effect on the reactivity of TFA isomers through reactions \ref{eq:3} and \ref{eq:4} (abstraction from the SH group). These two reactions, in turn, are the fastest from the present work. Reaction  \ref{eq:3} happens only scarcely, due to the SH group partaking in binding with the surface (0.88 of the molecules are locked, see Table \ref{tab:locTFA}). On the other hand, the localization fraction of t-TFA, of 0.63, suggests preferential t-TFA destruction. This finding contrasts with the chemical behavior found when comparing reactions \ref{eq:5} and \ref{eq:6}. In this case, the rate constants vary by $\sim$2 orders of magnitude with a negligible effect of the orientation effects (both isomers form hardly any localized bond with the surface, see Table \ref{tab:locTFA}, column 4), suggesting preferential c-TFA destruction. Therefore, it is tough for us to gauge the specific importance of the two factors for TFA, and we indicate that tailored kinetic simulations are required to shed light on this topic. Moreover, all the reactions with TFA are very fast and competitive with H diffusion on the grains \citep{Hama2012b}. 

The abstraction of \ce{H} from any of the TFA isomers and positions combinations produces radicals that can further react with a specific stereochemistry \citep{Molpeceres2021d}. Reactions \ref{eq:5} and \ref{eq:6} form either c/t-OCSH. While c/t-OCSH react with H to produce t/c-TFA (and sometimes \ce{H2S} and \ce{CO} \citep{Nguyen2021, molpeceres2021diastereoselective}), we found in our previous work that OCHS (produced from reaction \ref{eq:3} or reaction \ref{eq:4}) reacting with H produces either t-TFA or c-TFA with an estimated branching ratio of 0.67/0.33. A destruction of TFA through either reaction \ref{eq:3} or reaction \ref{eq:4} will eventually produce an enrichment on t-TFA. Nonetheless, one of the conclusions of our previous work was that OCS hydrogenation will produce hardly any c-TFA. With these new results we show, that even though it will still remain a small fraction, hydrogenation reactions on the grain can explain a small abundance of c-TFA, suggested by the recent tentative detection by us \citep{delaconcepcion2021transcis}. 

The case of FA is more straightforward. The reactivity of FA with H in our considered reactions is so different between isomers that, according to our results, no c-FA should be present on the surface of interstellar dust grains. Besides, we have not found evidence of localization effects of the CH moiety on FA. The presence of c-FA in cold regions (found in a fraction of $\sim$ 1:20 to t-FA \citep{Taquet2017, Agundez2019, delaconcepcion2021transcis} still remains a mystery. Indeed \cite{delaconcepcion2021transcis} have recently found that the isomerization from c-FA to t-FA should dominate and therefore, no c-FA should be present in the gas phase of cold clouds. Similarly, in this work, we have found the same behavior, this time on grains. In summary, our results show that c-FA should not withstand the conditions of the cold ISM, and yet it does. Therefore, we continue investigating the presence of c-FA in the ISM under various hypotheses, such as preferential gas-phase reactivity and photochemistry in the gas phase summed to the influence of \emph{non-thermal} effects on the surface of the dust grains.

Parallel to \ce{H} abstraction reactions, H additions might occur, affecting the overall balance of cis and trans isomers of both acids. In the case of FA, \cite{Chaabouni2020} reports that \ce{CO2} and \ce{H2O} are the main products of the addition reactions on ASW. Reactions on other substrates (pyrolitic graphite) also lead to other organics, e.g., \ce{H2CO} and \ce{H3COH} but their presence is minor on ASW. As was mentioned in the introduction, the same reaction was neither observed in other experiments \citep{Bisschop2007} nor in the present study. Independent of whether the reaction is viable or not, a destruction channel forming \ce{CO2} and \ce{H2O} (or \ce{CO2} and \ce{SH2} in the case of TFA) is irreversible and does not contribute to isomerization processes, that are the focus of the present work. In the light of our results, t-FA, c-TFA, and t-TFA could be found on dust grains.  

\subsection{Deuterium Fractionation of FA and TFA} \label{sec:deuterium}

In addition to providing insight into the isomerism of both FA and TFA, the reactivity of the different species involved in this study with deuterium is important to elucidate fractionation effects. Two possibilities are at play. First, and after formation of \ce{H2}, deuterium addition can proceed. Second, we can have \ce{HD} formation, where an incoming deuterium atom is the one reacting with the molecules. 

For the D addition reactions, the same discussion as above for isomerism holds, and therefore we hypothesize that for TFA, all single deuterated species (\ce{t-HCOSD}, \ce{c-HCOSD}, \ce{t-DCOSH}, \ce{c-DCOSH}) are plausible. Similarly, double deuterated species are likely to be present. The computational cost of computing the instanton rate constants for all possible reaction/site combinations is very high. We, therefore, estimate that based on the high rate constants for \ce{HD} abstraction (see below), the equivalent reactions from monodeuterated TFA are equally fast. For FA, only \ce{t-DCOOH} is expected, owing to the endothermicity of the \ce{H} abstraction from the OH group and the slow kinetics of reaction \ref{eq:2}. Interestingly, and following this reasoning, the presence of \ce{t-DCOOH} may imply a conversion on grains, indicating that at least some c-FA should deplete on top of them, where it is converted in t-FA. In Section$\,$\ref{deuterated_obs}, we will explore whether  \ce{t-DCOOH} is detected under the coldest conditions of the interstellar medium as those found in pre-stellar cores.

\begin{figure}
    \centering
        \includegraphics[width=\linewidth]{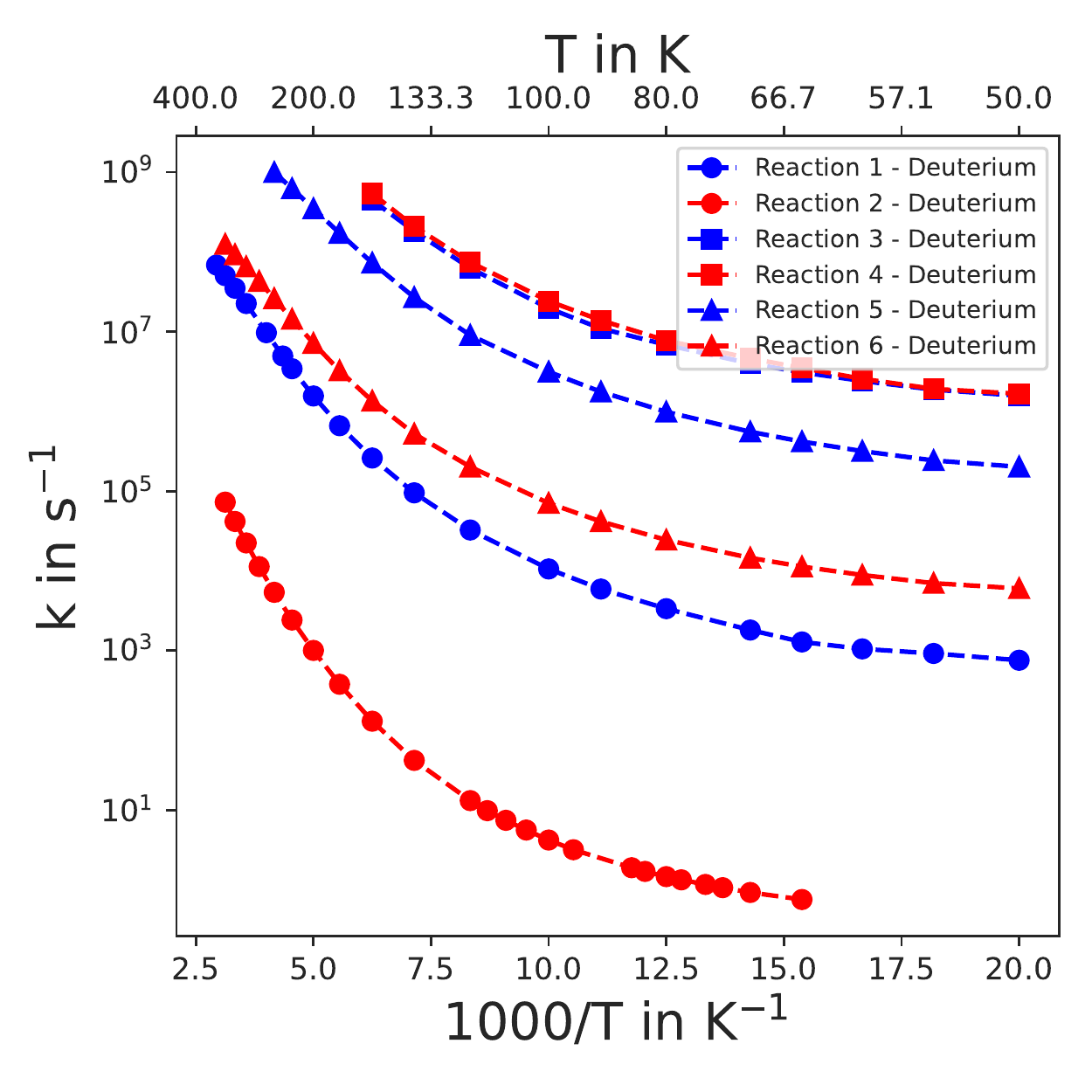} \\
    \caption{Rate constants for all the \ce{HD} abstraction reactions considered in the present work.}
    \label{fig:DRates}
\end{figure}

Finally, we have also estimated the rate constants for \ce{H} abstractions with deuterium with the deuterated counterpart reactions of reactions \ref{eq:1} to \ref{eq:6}. All of them, except for reaction \ref{eq:2} proceed fast enough, holding a very similar discussion as for the \ce{H} abstraction reactions and showing KIE of around $\sim$ 10--20. In the case of reaction \ref{eq:2} and due to the smaller importance of tunneling at low temperatures, the reaction presents an inverse KIE of $\sim$ 0.5. This occurs due to the reduced magnitude of the zero-point energy in the vibrational adiabatic barrier, rendering an overall lower barrier (by 4 kJ mol$^{-1}$) that surpasses the effect of quantum tunneling at low temperatures. Whether there is a real inverse KIE or a null one will require computational methods capable of reaching spectroscopic accuracy. However, as we showed before, the reaction is not competitive to hydrogen (or deuterium) diffusion, and therefore, such an effect is uninteresting from an astrochemical point of view. The specific KIE for each reaction are summarized in Table \ref{tab:KIE} and the rate constants and corresponding fits are depicted in Figure \ref{fig:DRates} and Table \ref{tab:FITD}. We have not observed a significant preference for the reactions' isomerism when using D atoms. Hence, all the trends discussed for \ce{H} abstraction reactions hold here for \ce{HD} ones.

To summarize, the reactivity of FA is clearly defined with preferential destruction of c-FA. The abstraction of H from c-FA forms t-HOCO that, after a subsequent H addition, reforms FA, with unknown isomerism. However, since t-FA does not experience such an abstraction, eventually all c-FA would be converted to t-FA, implying an excess of t-FA on grains by \ce{H} abstraction reactions. This also holds for deuterations. In contrast, TFA is very reactive in all H-bearing positions, with some reactions showing preferential destruction thanks to either fast overall reaction or specific binding with the surface. There exists a really complex interplay of \ce{H2}/\ce{HD} abstractions and \ce{H}/\ce{D} additions that requires a careful inspection by tailored kinetic models.

\begin{table}
\caption{Kinetic isotope effects (KIE) at the lowest temperature ($T$ in K) for the \ce{H} abstraction reactions and their HD counterparts.}             
\label{tab:KIE}      
\centering                          
\begin{tabular}{c c c}        
\hline\hline                 
Reaction & $T$ & KIE \\    
\hline                        
Reaction \ref{eq:1} & 50 & 13.6 \\
Reaction \ref{eq:2} & 65 & 0.51 \\
Reaction \ref{eq:3} & 50 & 22.4 \\
Reaction \ref{eq:4} & 50 & 21.7 \\
Reaction \ref{eq:5} & 50 & 21.7 \\
Reaction \ref{eq:6} & 50 & 16.1 \\
\hline                                   
\end{tabular}
\end{table}

\begin{table}
\caption{Same as Table \ref{tab:FITH} but for HD abstraction reactions.}             
\label{tab:FITD} 
\resizebox{\linewidth}{!}{
\centering                          
\begin{tabular}{c c c c c}        
\hline\hline                 
Reaction & $\alpha$ (s$^{-1}$) & $\beta$ & $\gamma$ (K) & $T_{0}$ (K) \\    
\hline                        
   Reaction \ref{eq:1}$^{a}$ (HD) & 1.25\e{10} & 1.70 & 1.59\e{3} & 1.32\e{2}\\
   Reaction \ref{eq:2} (HD) & 1.03\e{6} & 5.18 & 1.26\e{3} & 1.24\e{2} \\
   Reaction \ref{eq:3} (HD) & 1.15\e{13} & -4.37 & 1.48\e{3} & 9.42\e{1} \\
   Reaction \ref{eq:4} (HD) & 6.93\e{10} & 6.86 & 5.72\e{2} & 1.09\e{2} \\
   Reaction \ref{eq:5}$^{b}$ (HD) & 1.26\e{14} & -7.85 & 2.12\e{3} & 1.05\e{2} \\
   Reaction \ref{eq:6}$^{a}$ (HD) &  1.32\e{12} & -4.08 & 2.02\e{3} & 1.21\e{2} \\
\hline                                   
\end{tabular}}
\tablefoot{$^{a}$: Best fit deviated up to a factor 4 at low temperatures. $^{b}$: Best fit deviated up to a factor 3 at low temperatures. }
\end{table}


\section{Comparison with observations of deuterated formic acid}
\label{deuterated_obs}

As reported in Section$\,$4.2, our kinetic calculations of the deuterium abstraction reactions for FA reveal that only trans-DCOOH is expected to be detected in the interstellar medium due to the endothermicity of the H$_2$ abstraction reactions from the OH group. For hot sources such as hot corinos or hot cores, the formation efficiency of trans-DCOOH and trans-HCOOD on dust grains increases with increasing temperature (see Figure$\,$11) and therefore, one would expect to have higher chances of detection for this form of deuterated formic acid. 

Consistently with our results, \citet[][]{jorgensen18} reported the detection of trans-DCOOH toward the hot corino in IRAS16293-2422 B. However, they also present the discovery of trans-HCOOD toward the same source with a similar abundance. These results do not necessarily contradict our findings because FA could also form in the gas phase \citep{skouteris18}. It is however unknown whether deuterated formic acid can form efficiently in the gas phase, although we note that this formation route has been found to be efficient for more complex organics such as formamide \citep[][]{skouteris17}. 

For cold sources such as starless or pre-stellar cores, Figure$\,$11 reveals that trans-DCOOH should again be the most abundant form of deuterated formic acid in these sources, if produced on dust grains. To our knowledge, there are no observations of deuterated formic acid previously reported toward this type of sources and thus, we have searched for it toward the L1544 pre-stellar core using the IRAM 30m high-sensitivity observations of \citet[][]{JimenezSerra2016} and the ASAI Large Program data of \citet[][]{lefloch17}. Spectroscopic information for both the cis and trans isomers of DCOOH and HCOOD is available from the Jet Propulsion Laboratory (JPL) molecular line catalogue \citep[see][]{pickett98}. 

To search for these species, we have used the SLIM (Spectral Line Identification and Modelling) tool within the MADCUBA package\footnote{Madrid Data Cube Analysis on ImageJ is a software developed at the Center of Astrobiology (CAB) in Madrid; https://cab.inta-csic.es/madcuba/} \citep[][]{martin19}. SLIM generates the synthetic spectra of a molecular species assuming Local Thermodynamic Equilibrium (LTE) and derives the best fit of the observed spectra by using the Levenberg-Marquardt algorithm using the AUTOFIT tool. The fit provides the physical parameters column density (N$_{obs}$), excitation temperature (T$_{ex}$), velocity (V$_{LSR}$), and linewidth ($\Delta_v$).

The L1544 datasets do not reveal any emission detected from either c/t-DCOOH or c/t-HCOOD. In Table$\,$\ref{tab:deuterium}, we report the rms noise level and the upper limits of the column density and molecular abundance derived for all isomers of deuterated formic acid. The rms noise level has been calculated by considering the spectra within a velocity range of 5$\,$km$\,$s$^{-1}$ around each of the lines shown in Table$\,$\ref{tab:deuterium}. The upper limits to the column density and abundance of all forms of deuterated formic acid have been calculated assuming a radial velocity of V$_{LSR}$=7.2$\,$km$\,$s$^{-1}$ and a linewidth of $\Delta_v$=0.4$\,$km$\,$s$^{-1}$. The derived column densities range from $\leq$2.9$\times$10$^{10}$ to $\leq$1.5$\times$10$^{11}$$\,$cm$^{-2}$, while the measured abundances range from $\leq$8.4$\times$10$^{-13}$ to $\leq$4.2$\times$10$^{-12}$.

Our results from Section$\,$4.2 indicate that trans-DCOOH could be present in the ISM since it is expected to form efficiently on the surface of dust grains. Although we have not detected trans-DCOOH toward L1544, this non-detection, however, does not allow us to rule out its grain surface formation because the derived abundance upper limit (of $\leq$2.0$\times$10$^{-12}$) is not stringent enough. Indeed, if we consider the abundance of t-HCOOH measured toward the position of L1544's dust peak \citep[of 8.9$\times$10$^{-12}$;][]{jimenez21}, and the D/H ratio determined for this source \citep[of 25\%;][]{redaelli19}, the expected trans-DCOOH abundance is $\sim$2.2$\times$10$^{-12}$, i.e. of the same order as the upper limit obtained for this molecule and source. Therefore, higher-sensitivity observations will be needed to establish whether trans-DCOOH is present in the cold gas of starless/pre-stellar cores.

\begin{table}
\caption{Upper limits of the column density of trans/cis-DCOOH and trans/cis-HCOOD measured toward the dust peak of L1544}             
\label{tab:deuterium}
\resizebox{\linewidth}{!}{
\centering                          
\begin{tabular}{lccccc}        
\hline\hline                 
Species & Transition & Frequency & rms & Column Density & Abundance \\
& & (MHz) & (mK) & (cm$^{-2}$) & \\ \hline
t-DCOOH & 4$_{0,4}$$\rightarrow$3$_{0,3}$ & 87340.8922 & 1.5 & $\leq$7.1$\times$10$^{10}$ & $\leq$2.0$\times$10$^{-12}$ \\
c-DCOOH & 4$_{0,4}$$\rightarrow$3$_{0,3}$ & 85616.8990 & 2.3 & $\leq$2.9$\times$10$^{10}$ & $\leq$8.4$\times$10$^{-13}$ \\
t-HCOOD & 4$_{0,4}$$\rightarrow$3$_{0,3}$ & 86493.0196 & 3.4 & $\leq$1.5$\times$10$^{11}$ & $\leq$4.2$\times$10$^{-12}$ \\
c-HCOOD & 4$_{0,4}$$\rightarrow$3$_{0,3}$ & 81872.8197 & 7.0 & $\leq$8.2$\times$10$^{10}$ & $\leq$2.3$\times$10$^{-12}$ \\
\hline
\end{tabular}}
\tablefoot{Molecular abundances are calculated using a H$_2$ column density of 3.5$\times$10$^{22}$\,cm$^{-2}$ for the dust continuum peak of L1544 see \cite{JimenezSerra2016}.}
\end{table}

\section{Conclusions}

We have studied the hydrogen abstraction reactions of characteristic small acids, formic acid, and thioformic acid under interstellar conditions finding an overall high reactivity of these archetypes and suggesting that the reactions should be relevant for the chemistry of the ISM under cold conditions. The main conclusions of the present work are therefore

   \begin{enumerate}
      \item Hydrogen abstraction reactions on formic acid occur on its cis isomer almost five orders of magnitude faster than the same reactions on the trans isomer. This finding, summed to the overall lack of reactivity of trans-formic acid, that was also confirmed by our experiments, heavily implies that no trans-cis interconversion may happen on the surface of ice-coated dust grains. Cis-trans interconversion can, in principle proceed, in the absence of competitive addition reactions.
      \item In contrast with formic acid, H atoms are abstracted from thioformic acid very fast. Abstractions at the thiol group are the fastest reactions found in this work and proceed at a similar rate for both isomers. Abstractions on the CH group are slower but present a marked dependence on the isomerism of the parent acid.
      \item We calculated the binding energies on amorphous solid water for all isomers of the acids considered here. Their values are about $-6000$~K in the case of the formic acid and $-3100$~K for thioformic acid. Both isomers (cis and trans) have almost the same binding energies. 
      \item The binding of the acids with ASW is highly localized owing to hydrogen bonds between the acid and the surface. We have analyzed to which extent these orientations inhibit reactivity, finding that they do not have any effect in the case of formic acid. However, they are important to predict the chemistry of thioformic acid, especially concerning the hydrogen abstractions at the thiol group. From our statistical sampling of binding sites, 88\% of cis-thioformic acid ones form a hydrogen bond with the surface, in contrast with the 63\% of trans-thioformic acid.
      \item We repeated our study for abstraction reactions by D atoms, finding equivalent conclusions as for \ce{H} and kinetic isotopic effects of around 10--20 except for trans-formic acid where the lower relevance of quantum tunneling yields an inverse KIE.
      \item Our recent tentative detection of cis-thioformic acid is supported by our calculations. The chemistry of thioformic acid on grains is very complicated, and our understanding of cis/trans ratios in the interstellar medium would benefit from extensive and specific modeling.
      \item The mechanisms considered here cannot explain the presence of a small amount of cis-formic acid in cold cores. This is in agreement with theoretical gas-phase calculations but in contrast with observations. Therefore, additional mechanisms must be at play to explain the presence of cis-formic acid in regions with a low density of photons such as molecular clouds.
      \item Both t-DCOOH and t-HCOOD has been reported toward the hot gas of the IRAS16293-2422B hot corino with similar abundances. This would seem to contradict our results. However, it is possible that deuterated formic acid also forms efficiently in the gas phase. Future kinetic calculations will be needed to establish this route as a viable formation mechanism. At the cold temperatures of the L1544 pre-stellar core, neither c/t-DCOOH nor c/t-HCOOD are detected, although their derived upper limits are not stringent enough to disentangle whether they are formed on grain surfaces or in the gas phase.
   \end{enumerate}

The here presented results are complemented with the data found in the literature for addition reactions \citep{Cao2014,Chaabouni2020}. Similar calculations on the addition reactions to TFA are a viable continuation to the present work, seeking an integral understanding of the chemistry of simple acids on top of interstellar dust grains.

\begin{acknowledgements}

Computer time was granted by the state of Baden-W\"urttemberg through bwHPC and the German Research Foundation (DFG) through grant no. INST 40/467-1FUGG is greatly acknowledged. G.M thanks the Alexander von Humboldt Foundation for a post-doctoral research grant. We thank the Deutsche Forschungsgemeinschaft (DFG, German Research Foundation) for supporting this work by funding EXC 2075 - 390740016 under Germany's Excellence Strategy. We acknowledge the support by the Stuttgart Center for Simulation Science (SimTech). I.J.-S. has received partial support from the Spanish State Research Agency (AEI) through project numbers PID2019-105552RB-C41 and MDM-2017-0737 Unidad de Excelencia “Mar\'ia de Maeztu”- Centro de Astrobiología (CSIC-INTA). We also acknowledge support from the Spanish National Research Council (CSIC) through the i-Link project number LINKA20353. This work is also supported by JSPS KAKENHI grant Nos. JP21F21319, JP21H04501, JP21H05414, JP21K18639, and JP17H0687
This work also made use of ASAI $"$Astrochemical Surveys At IRAM$"$ \citep[][]{lefloch17}.

\end{acknowledgements}

%
%

\bibliographystyle{aa}
\bibliography{sample.bib}

\end{document}